\documentclass[12pt,preprint]{aastex}
\usepackage{epsfig}

\def\vkm{km s$^{-1}$}
\def\Jyb{Jy Beam$^{-1}$}
\def\Jybk{Jy beam$^{-1}$ km s$^{-1}$}

\def\arcs#1{$#1''$}

\def\ra#1#2#3#4{#1^\mathrm{h} #2^\mathrm{m} #3^\mathrm{s}_{^\textrm{.}} #4}
\def\dec#1#2#3#4{#1\degr #2\arcmin #3^{\prime\prime}_{^\textrm{.}}#4}
  
\def\H2{H$_2$}

\def\vsys{V_\textrm{\scriptsize sys}}

\def\solarM{M_\odot}

\def\rhoH{\rho_\textrm{\scriptsize H$_2$}}

\def\Tgas{T_\textrm{\scriptsize gas}}
\def\Tin{T_\textrm{\scriptsize in}}
\def\FT{F_\textrm{\scriptsize T}}
\def\Rin{R_\textrm{\scriptsize in}}
\def\Rout{R_\textrm{\scriptsize out}}
\def\Va{V_\textrm{\scriptsize a}}
\def\Vb{V_\textrm{\scriptsize b}}
\def\Vx{V_\textrm{\scriptsize x}}
\def\Vy{V_\textrm{\scriptsize y}}
\def\Vz{V_\textrm{\scriptsize z}}

\def\Vturb{V_\textrm{\scriptsize turb}}
\def\Vth{V_\textrm{\scriptsize th}}

\def\tsw{t_\textrm{\scriptsize sw}}
\def\tdyn{t_\textrm{\scriptsize dyn}}
\def\Rsw{R_\textrm{\scriptsize sw}}

\def\tagb{t_\textrm{\scriptsize agb}}

\def\swMlos{\dot{M}_\textrm{\scriptsize SW}}
\def\agbMlos{\dot{M}_\textrm{\scriptsize AGB}}
\def\eqMlos{\dot{M}_\textrm{\scriptsize eq}}

\begin{document}

\title{Remodel the Envelope Around the 21 $\mu$m PPN IRAS 07134+1005}
\author{Chun-Hui Yang\altaffilmark{1,2} and Chin-Fei Lee\altaffilmark{1}}
\altaffiltext{1}{Academia Sinica Institute of Astronomy and Astrophysics, P.O. Box 23-141, Taipei 106, Taiwan}
\altaffiltext{2}{Department of Physics, National Taiwan Normal University, 88, Sec.4, Ting-Chou Rd., Taipei 116, Taiwan; chyang@asiaa.sinica.edu.tw}

\begin{abstract}
Recently, the CO J=3-2 observational result of the envelope of the 21
$\mu$m PPN IRAS 07134+1005 has been reported. Assuming that the CO
J=3-2 line was optically thin, the mass-loss rate of the superwind in this
PPN was found to be at least 2 orders of magnitude lower than the typical
range. In order to obtain a more accurate mass-loss rate, we reexamine this
data and construct a radiative transfer model to compare with the data. 
Also, in order to better resolve the superwind, we adopt a different
weighting on the data to obtain maps at higher resolution. Our result shows
that the CO J=3-2 emission is located slightly further away from the central
source than the mid-IR emission, probably because that the material is
cooler in the outer part and thus better traced by the CO emission. At lower
resolution, however, the CO emission appeared to be spatially coincident
with the mid-IR emission. Our model has two components, an inner ellipsoidal
shell-like superwind with an equatorial density enhancement and an outer
spheroidal AGB wind. The thick torus in previous model could be considered
as the dense equatorial part of our ellipsoidal superwind. With radiative
transfer, our model reproduces more observed features than previous model
and obtains an averaged superwind mass-loss rate of $\sim$ 1.8 $\times\
10^{-5}\ \solarM$ yr$^{-1}$, which is typical for a superwind.  The
mass-loss rate in the equatorial plane is 3 $\times\ 10^{-5}\ \solarM$
yr$^{-1}$, also the same as that derived before from modeling CO J=1-0
emission.
\end{abstract}


\section{Introduction}
Proto-planetary nebula (PPN) is an object in a transient phase between an
asymptotic giant branch (AGB) phase and a planetary nebula phase in the end
stage of a low to intermediate mass (1$-8 \solarM$) star. It consists of a
post-AGB stellar core (star) and an extensive circumstellar envelope of dust
and gas. It is bright at infrared (IR) wavelength because of the dense and
warm circumstellar dust. Recently, a large number of observations have
revealed that PPNe are mostly bipolar, multipolar or elliptical with a
torus-like structure in the equator. Moreover, most of them have an
extensive and roughly round halo, produced by the progenitor AGB star (see,
e.g., review by Balick \& Frank 2002), indicating that there is a structural
change from AGB to PPN phase. The shaping mechanism of PPNe is still
unclear and believed to be closely related to the mass-loss process of the
post-AGB star.

IRAS 07134+1005 (hereafter I07134) is a well-studied carbon-rich PPN with a
21 $\mu$m feature \citep{kwok 1989, volk 1999} that is also seen in some
other carbon-rich PPNe. The chemical study of I07134 reveals a metal-poor
central star with abundant C, N, O, and s-process elements, indicating that
the star has gone through third dredge-up when it was on the AGB phase
\citep{van winckel and reyniers 2000}. \cite{barthes 2000} studied the
pulsation of the central star revealing a variation of 36.8 days, indicating
a stellar mass of 0.6 $\solarM$.

In IR images \citep{kwok 2002,ueta 2005}, I07134 was seen with an elliptical dusty shell with an equatorial density enhancement (i.e. a torus-like structure) surrounding its central star. In the optical, the elliptical shell was seen surrounded by a roughly round halo produced by the AGB wind in the past \citep{ueta 2000}. The CO J=1-0 map \cite[][hereafter M2004]{meixner 2004} also showed a torus-like structure similar to that seen in the mid-IR image. A radiative transfer model made by M2004 indicates that the envelope of I07134 
has two components, an inner warm and dense superwind corresponding to the elliptical shell, and an outer cool and sparse AGB wind corresponding to the round halo.

\citet[hereafter N2009]{nakashima 2009} have reported the CO J=3-2 observation of this PPN obtained with the Submillimeter Array [SMA \citep{ho 2004}]. Unlike CO J=1-0, CO J=3-2 traces mainly the superwind component because it traces warmer and denser material than CO J=1-0. N2009 have proposed a morpho-kinematics model to compare with the observation. They assumed the CO J=3-2 line is optically thin and estimated a lower limit of the mass-loss rate, which is at least two orders of magnitude lower than the typical range (10$^{-7}$ to 10$^{-4}$ $\solarM$ yr$^{-1}$, see, e.g., the review by Van Winckel 2003). For studying the shaping mechanism of I07134, we need to know the mass-loss rate, spatial structure and kinematics of the superwind component as accurate as possible. Therefore, in this paper, we reexamine the CO data and construct a radiative transfer model to compare with the data. The details about our data reduction and mapping are described in Section 2. The observation results are presented in Section 3. Our model is described and compared with the observation in Section 4. We discuss and conclude our work in Sections 5 and 6, respectively.

\section{Observation}
The SMA CO J=3-2 observation of I07134 in this paper has been reported by N2009 and please refer to their paper for the details. In this Section, we only summarize the parameters of this observation in Table \ref{tab:obs}, and describe the differences between our and their data reduction and mapping.

8 antennas were used in the observation. We adopted the data from 7 (one more than N2009) of them in data reduction and mapping because antenna 7 was excluded due to its weak fringes and amplitudes over the whole observation. In addition, we removed the first half data of antenna 4 because of its scattering phase. In mapping, a ``super-uniform'' weighting is used to achieve the best compromise between the sensitivity and angular resolution. It results in an angular resolution of \arcs{1.66}$\times$\arcs{1.46} with a position angle (PA) of 17$^\circ$, which is higher by 40\% in beam area than that in N2009, showing the compact structure clearer. Our maps will show similar features to those of N2009 if we convolve our maps with their angular resolution. The channel maps have a resolution of $\sim$ 0.7 \vkm{} per channel, with a rms noise (hereafter $\sigma$) of 0.27 \Jyb. They are used to produce integrated intensity map, position velocity (PV) diagrams, and spectrum. The channel maps we shown in this paper are binned to have a lower resolution of $\sim$1.4 \vkm{} per channel to show how the structure changes with velocity. Although this velocity resolution is lower by 40\% than that of N2009, it is enough to show the main structure of I07134 in different velocity.

\section{Observational Results} 
    \subsection{Integrated intensity map}
\begin{figure}{}
\begin{center}
\includegraphics[width=165mm]{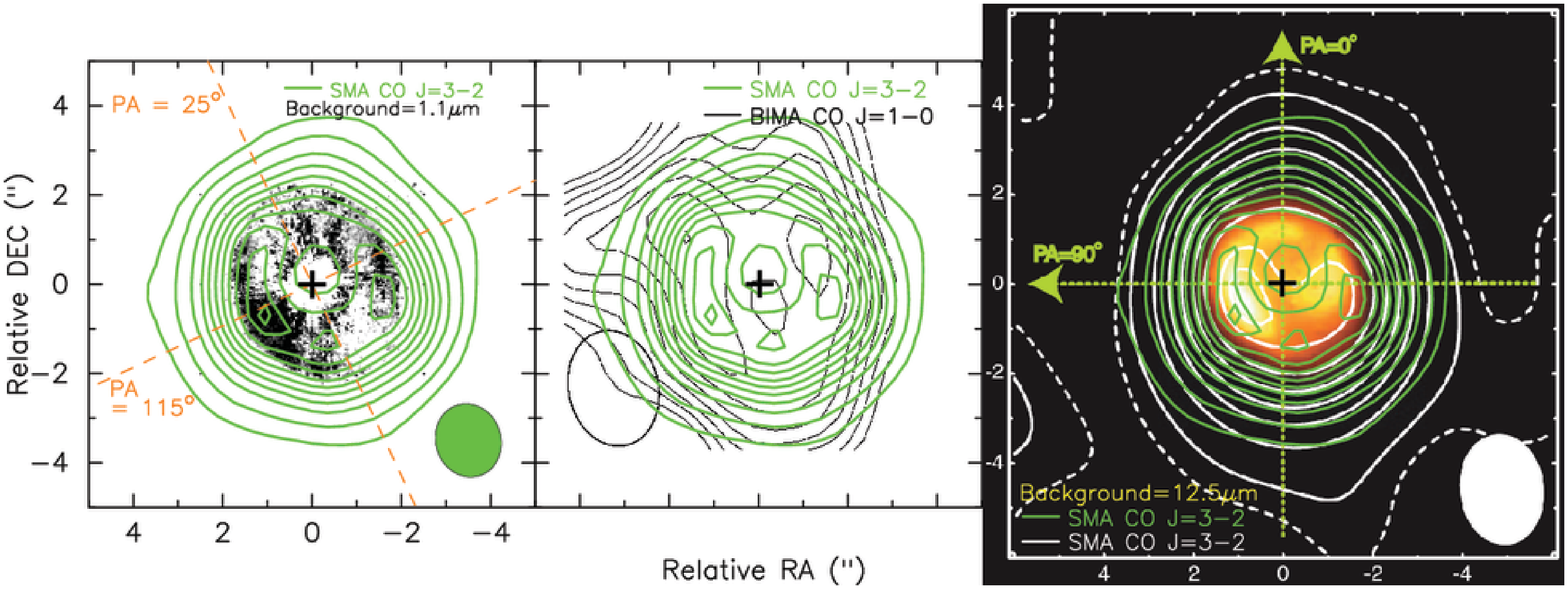}
\caption[Integrated intensity maps]
{{Integrated intensity map of the SMA CO J=3-2 emission (green contours) superposed on (\textsl{left}) the near-IR (1.1$\mu$m) image \citep{ueta 2005}, (\textsl{center}) the CO J=1-0 map \citep{meixner 2004}, (\textsl{right}) the integrated
CO J=3-2 intensity map of N2009 (white contours) and the mid-IR (12.5$\mu$m) image [color background \citep{kwok 2002}], respectively, in the same scale. The integrated intensity map is integrated from 57.6 to 86.5 \vkm{}. The ``+" symbol marks the position of the central star. The first and last contours of the CO J=3-2 emission are 2.8$\sigma$ and 66.2$\sigma$, respectively, and the middle contours are from 7$\sigma$ to 63$\sigma$ with a step of 7$\sigma$, where 1$\sigma$ is 1.23 \Jybk. The first contour level is the same as that of the integrated map of N2009. The resolutions of the CO J=3-2 emission of this work (\arcs{1.66}$\times$\arcs{1.46}, PA$=17^\circ$) and of N2009 are shown in
the bottom-right corner of (\textsl{left}) and (\textsl{right}), respectively, and the resolution of CO J=1-0 emission is shown in bottom-left corner of (\textsl{center}).}\label{co32_nearIR_midIR_co10}}

\end{center}
\end{figure}

In Figure \ref{co32_nearIR_midIR_co10}, we superposed our SMA CO J=3-2 (integrated intensity) map on the near-IR \citep{ueta 2005}, CO J=1-0 (M2004), and mid-IR \citep{kwok 2002} maps as well as the CO J=3-2 map of N2009 to analyze the structure of I07134. The near-IR emission mostly traces the light of the central star scattered by the surrounding dust, and the mid-IR emission mostly traces the thermal dust emission. The near-IR map shows a well-defined elliptical shell with a major axis at PA $\sim$ 25$^\circ$ \citep{ueta 2005}, and the mid-IR map shows a roughly round envelope surrounding the central star. The maps all show a double-peak structure around the minor axis (PA $\sim$ 115$^\circ$), indicating that the circumstellar envelope of I07134 has an equatorial density enhancement (i.e., a torus-like structure). At low resolution, the two peaks of the CO J=3-2 emission appeared to be spatially coincident with those in the mid-IR \citep{nakashima 2009}. But here at higher resolution, the two peaks of the CO emission are located slightly further away from the source than those in the mid-IR, especially for the western peak. The two peaks are asymmetric about the central star, with the eastern peak closer to the central star than the western peak by about 20\%.  Note that both the CO J=3-2 and the CO J=1-0 emission are weak in the north of the central star than that in the south, in contrast to that seen in the mid-IR.

    \subsection{Channel maps}
\begin{figure}[h]
\begin{center}
\includegraphics[width=120mm,angle=-90]{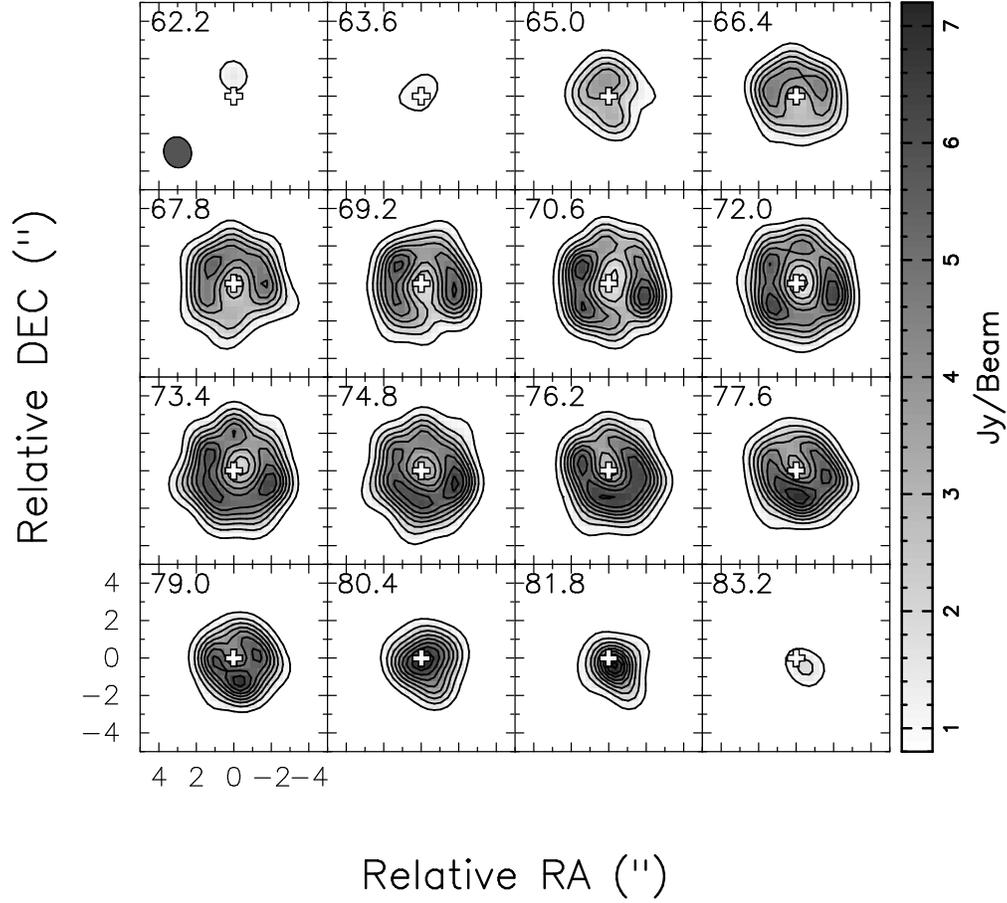}
\caption[Observed channel maps]
{{
Observed channel maps in contours and grey scale. The contour levels are from 4.3$\sigma$ to 39$\sigma$ with a step of 4.3$\sigma$, where $\sigma$ is 0.19 \Jyb. Top-left corner of each channel map shows the LSR velocity in \vkm. The systemic velocity is in the channel of 72 \vkm. The velocity width is $\sim$ 1.4 \vkm. 
The  white cross symbol at the center of each channel map marks the position of the central star. The resolution is shown in the bottom-left corner of the first channel map. 
The wedge shows the intensity scale in \Jyb.}\label{obs_cha}}
\end{center}
\end{figure}

Our channel maps (Figure \ref{obs_cha}) are similar to those of N2009 (their Figure 3), showing similar variation of the morphology with the velocity. For example, the size of the emission gradually increases from high blue- and red-shifted velocities toward the systemic velocity (72 \vkm, as found in our model described later), single peaks are seen at high blue- and red-shifted velocities, an opening (less emission) toward the south (i.e., inverted U-shape, ``convex upward" in N2009) from 66.4 to 69.2 \vkm, an opening toward the north (i.e., U-shape, ``convex downward" in N2009) from 73.4 to 79 \vkm, and two elongated and clumpy peaks in the east and west near the systemic velocity (from 70.6 to 72 \vkm). Notice that, the high velocity emission peaks are not exactly at the map center (central star position). The highest blue-shifted peak is shifted slightly to the north, and the highest red-shifted peak is shifted slightly to the south-west.

    \subsection{PV diagrams}
\begin{figure}[h]
\begin{center}
\includegraphics[width=60mm,angle=-90]{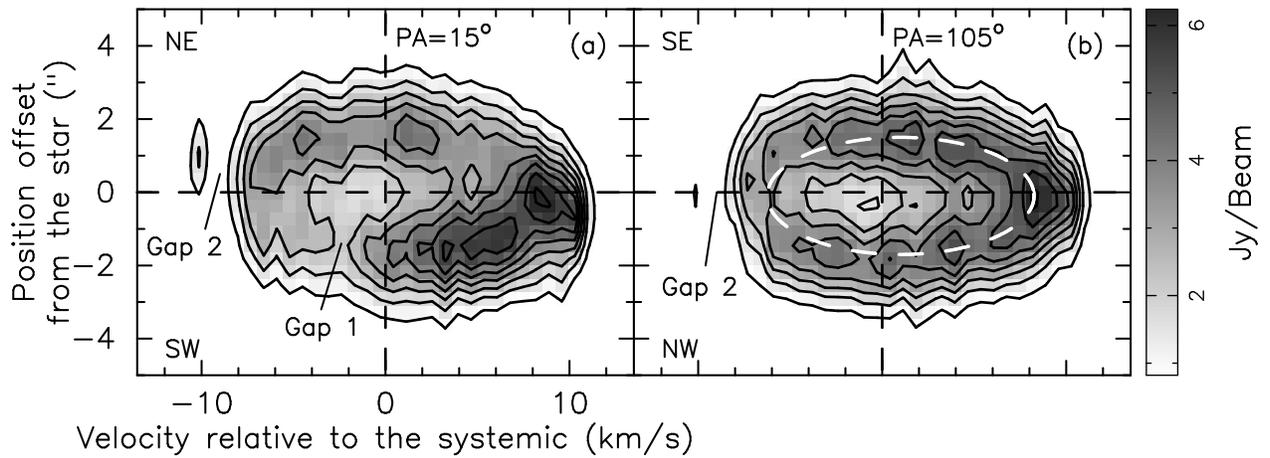}
\caption[Observed PV diagrams]
{{
Observed PV diagrams cut along (a) PA$=15^\circ$ and (b) PA $=105^\circ$, respectively, in contours and grey scale. The contour levels are from 3$\sigma$ to 23$\sigma$ with a step of 3$\sigma$, where $\sigma$ is 0.27 \Jyb. The capital NE, SW, SE and NW indicate the north-east, south-west, south-east and north-west directions on the observed integrated intensity map. The wedge shows the intensity scale in \Jyb.}\label{obs_pv}}
\end{center}
\end{figure}

The PV diagrams cut along the major (the poles, PA$=15^\circ$, as found in our model described later) and the minor (the equator, PA$=105^\circ$) axes of the ellipse are presented in Figure \ref{obs_pv}. Note that the major and minor axes here are different from those found in \cite{ueta 2005}. These PV diagrams are similar to those of N2009 that cut along the similar PAs (0$^\circ$ and 90$^\circ$, their Figure 4). At higher angular resolution, however, our PV diagrams show a clearer ring-like structure [as indicated by a white dashed ellipse on Figure \ref{obs_pv} (b)] in both cuts. The lack of emission inside the ring-like structure is consistent with a detached envelope around the central star. In the PV diagrams, there are two obvious asymmetric emission distributions. One is that the high red-shifted emission is stronger than the high blue-shifted emission toward the star position in both PV diagrams. The other is that in the PV diagram cut along the major axis, the south-western emission in the red-shifted part is much stronger than that in the blue-shifted part, producing a gap (Gap 1) there. There is another gap near the highest blue-shifted velocity (Gap 2), separating the highest blue-shifted emission from the main emission structure. Note that in the PV diagram cut along the major axis, the emission at the highest blue-shifted and red-shifted emission are not at the source position, consistent with that seen in the channel maps as mentioned above.

From the observation, we suggest that the envelope of I07134 is a radially expanding ellipsoidal shell, with the red-shifted emission stronger than the blue-shifted emission. This envelope is inclined to the plane of the sky, producing the U- and inverted U-shape structures in the channel maps. In the following section, we introduce our radiative transfer model and compare it with this observation in order to obtain more accurate mass-loss rate, morphology and kinematics of the envelope.

\section{An Expanding ellipsoidal model}
      \subsection{Model}        
\begin{figure}[h]
\begin{center}
\includegraphics[width=100mm,angle=-90]{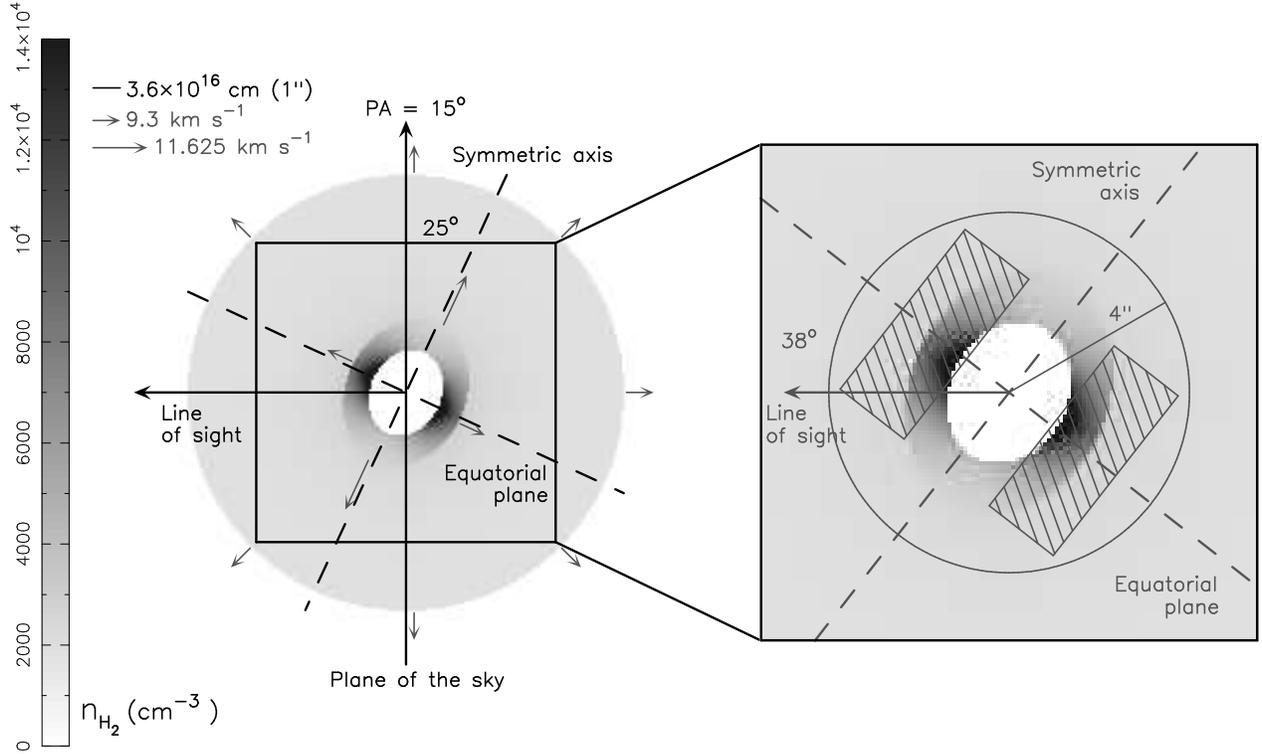}
\caption[The sketch of our model]
{{Left: A transect map of the density (grey scale) and velocity 
(grey arrows) distributions of the molecular envelope in our model (Section 4). This map is taken from a plane along the symmetric axis (PA=15$^\circ$). The inner dense elliptical region is the superwind component and the outer sparse spherical region is the AGB wind component. Right: A transect map of the model of N2009 on top of ours in the same scale but taken from a different PA, theirs is 8$^\circ$ and ours is 15$^\circ$. 
The grey circle shows the outer boundary of their spheroidal AGB wind 
and the two hatched rectangular regions show their thick torus-like superwind. In the right sketch, the axes and labels are for the model of N2009.}\label{model_sketch}}
\end{center}
\end{figure}

Here we construct a code with a model different from that of N2009 and include the radiative transfer to calculate the CO J=3-2 emission, rather than assuming an optically thin emission. As in M2004, the envelope has two components, an outer spherical AGB wind and an inner ellipsoidal shell-like superwind (Figure \ref{model_sketch}).  In the superwind component, we do not adopt a spheroidal shell as in M2004 nor a thick torus as in N2009, because the near-IR image [Figure \ref{co32_nearIR_midIR_co10} (a)] shows an elliptical shell-like morphology in the inner part of this envelope.  Thus, the superwind is assumed to be an ellipsoidal shell elongated in the north-south direction and looks like a football. In the cartesian coordinate system, the ellipsoidal shell can be described with the following equation,
\begin{equation}
     \frac{x^2}{a^2}+\frac{y^2}{ a^2}+\frac{z^2}{b^2} = 1,
\end{equation}
with the $x-y$ plane being the equatorial plane and the $z-$axis being the major axis in the north-south direction. Here, $a$ and $b$ are the one-half of the minor and major axes of the ellipsoidal shell with an constant ellipticity defined as $\epsilon=1-\frac{a}{b}$. The ellipsoidal shell has an inner radius $\Rin$ and a thickness $\Delta a$ in the minor axis, and thus $a=\Rin$ to $\Rin+\Delta a$ (which is the outer radius). On the other hand, the spherical AGB wind has an outer radius of $\Rout$, and its inner boundary is the outer boundary of the superwind.

We assume that the envelope is expanding radially, and the expansion
velocity of the superwind follows the same equation as the ellipsoidal shell
for self-consistency,
\begin{equation}
     \frac{\Vx^2}{\Va^2}+\frac{\Vy^2}{\Va^2}+\frac{\Vz^2}{\Vb^2} =1,
\end{equation} 
with $\Va$ and $\Vb$ being the expansion velocities in the minor (equator) and major (pole) axes, respectively. As can be seen from the equation, the expansion velocity increases from the equator to the poles, qualitatively similar to that expected in some of the ISW simulations \cite[e.g.,][]{dwarkadas 1996}. As for the AGB wind, it expands radially with a constant velocity of $\Va$.
 
\begin{figure}[h]
\begin{center}
\includegraphics[width=120mm]{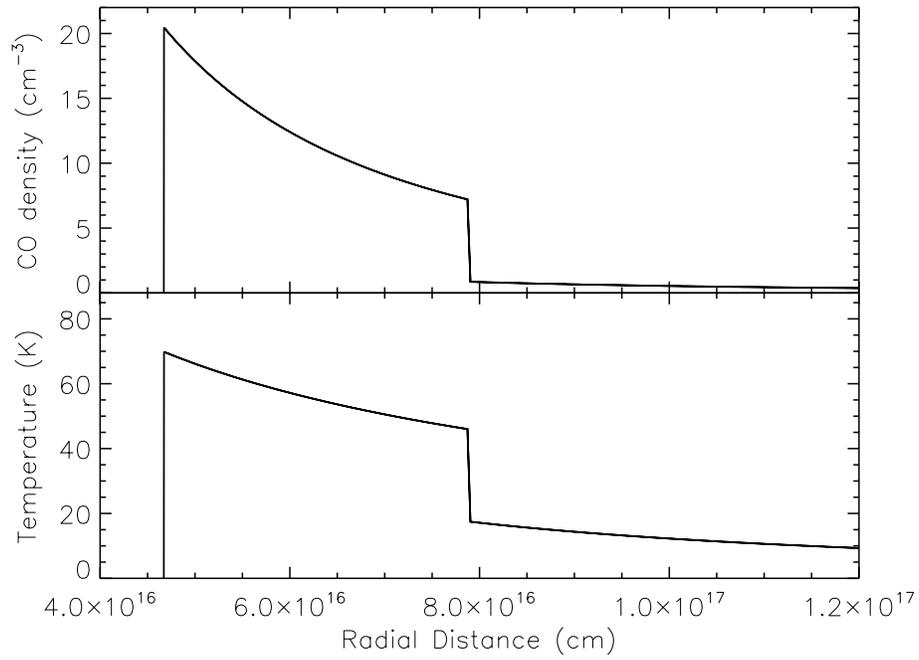}
\caption[density and temperature profiles]
{{Equatorial density (upper panel) and temperature (lower panel) of the molecular gas
as a function of radius for the superwind and AGB wind components in our model. 
Please note that we do not show the entire AGB wind region, which has an 
outer radius of $\sim 3 \times 10^{17}$ cm.
}\label{profiles}}
\end{center}
\end{figure}

Below are our density and temperature profiles of the molecular gas 
(molecular hydrogen) in the envelope (in spherical coordinate system).
\[ \rhoH\left(R,\theta\right) = \left\{
    \begin{array}{ll}
         \frac{\eqMlos}{4\pi R^2 \Va}f(\theta)   &   \mbox{Superwind component,} \\ 
         F\frac{\eqMlos}{4\pi R^2 \Va}  &    \mbox{AGB wind component,} 
    \end{array} 
\right . \]    
\[ \Tgas\left(R\right) = \left\{
    \begin{array}{ll}
        \Tin\left(\frac{R}{\Rin}\right)^\beta    &    \mbox{Superwind component,} \\
        \Tin \FT\left(\frac{R}{\Rsw}\right)^\gamma    &   \mbox{AGB wind component,} 
    \end{array} 
\right . \]    
where $\eqMlos$ is the mass-loss rate of the superwind component in the
equator. The density decreases with radius $R$ with a power-law index of -2, in both the superwind and AGB wind components, and it has a sudden drop with a factor $F$ at the superwind-AGB wind boundary. In order to reproduce the observed equatorial density enhancement, the density of the superwind component is multiplied by a
simple torus function 
\begin{equation}
f(\theta)=1-\alpha \cos\theta,  
\end{equation}
where $\theta$ is the angle from the $z$-axis, and the value of $\alpha$ is between 0 and 1. Here $f(\theta)$ equals to 1 at the equator and 1-$\alpha$ at the poles.
$\Tin$ is the temperature at $\Rin$. The temperature is assumed to decrease 
with $R$ with a power-law index of $\beta$ 
in the superwind and $\gamma$ in the AGB wind. 
It also has a sudden drop with a factor $\FT$ in the superwind-AGB wind boundary.
In the AGB wind, the temperature is set to 5 K when it decreases to below 5 K. The density and temperature profiles along the minor axis of the two components are shown in Figure \ref{profiles}.

Radiative transfer is used to derive the CO J=3-2 emission in our model. Thermal line width $\Vth$ and the line width due to the turbulence velocity $\Vturb$ are also included. The systemic velocity $\vsys$ is assumed to be a free parameter. Also, we rotate our model counterclockwise by a position angle (PA) from the $z$-axis and tilt it with an inclination angle $i$ with the north part tilted away from us. To compare the model results with the observation properly, we use MIRIAD \citep{miriad} to derive the model visibility from the model data cubes with the observed \textsl{uv}$-$coverage, and then use the same imaging procedure as we did for the observed channel maps to derive the model channel maps. As a result, the model and the observed channel maps have the same spatial and velocity resolutions. With the channel maps, we can obtain the integrated map, PV
diagrams, and spectrum of our model.

    \subsection{Model Results}

The best-fit parameters are listed in Table \ref{tab:mod}. There are 18
parameters in our model, 3 are constant and 15 are free. The 3 constant
parameters are distance of the source, CO abundance (CO/H$_2$ in number
density) and $R_\textrm{out}$. For the CO abundance, we adopt the value of
9.2 $\times$ 10$^{-4}$ from M2004. If we adopt the value of 7.4
$\times$ 10$^{-4}$ as in N2009, which is lower by
$\sim$ 20\% than that in M2004,
our mass-loss rate will increase only by $\sim$ 20\%. The 15 free
parameters are $\Rin$, $\Delta a$, $\epsilon$, PA, $i$, $\vsys$, $\Va$,
$\Vturb$, $\Tin$, $\FT$, {\bf $\beta$, $\gamma$}, $F$, $\eqMlos$ and
$\alpha$. We compare the results of our model with the observation, and
determine the best fit parameters and their error bars based on the
following five criteria. (1) The total flux of our model can not be more
or less by 10\% than that of the observation. Here the total flux is derived
from within a \arcs{4} wide box centered at the central star, where the
emission mainly comes from the superwind region. (2) The blue-shifted peak
of the spectrum of our model can not be more or less by 10\% than that of
the observation. (3) The highest contour in the moment map of our model can
not be more or less by one contour level than that of the observation. (4)
The lowest contour in the moment map of our model should be in between the
lowest and penultimate-low contours in the moment map of the observation.
(5) The morphology trend in the channel maps and PV diagrams in our model
should be similar to that in the observation as mentioned in Section 3.

\begin{figure}[h]
\begin{center}
\includegraphics[width=120mm,angle=-90]{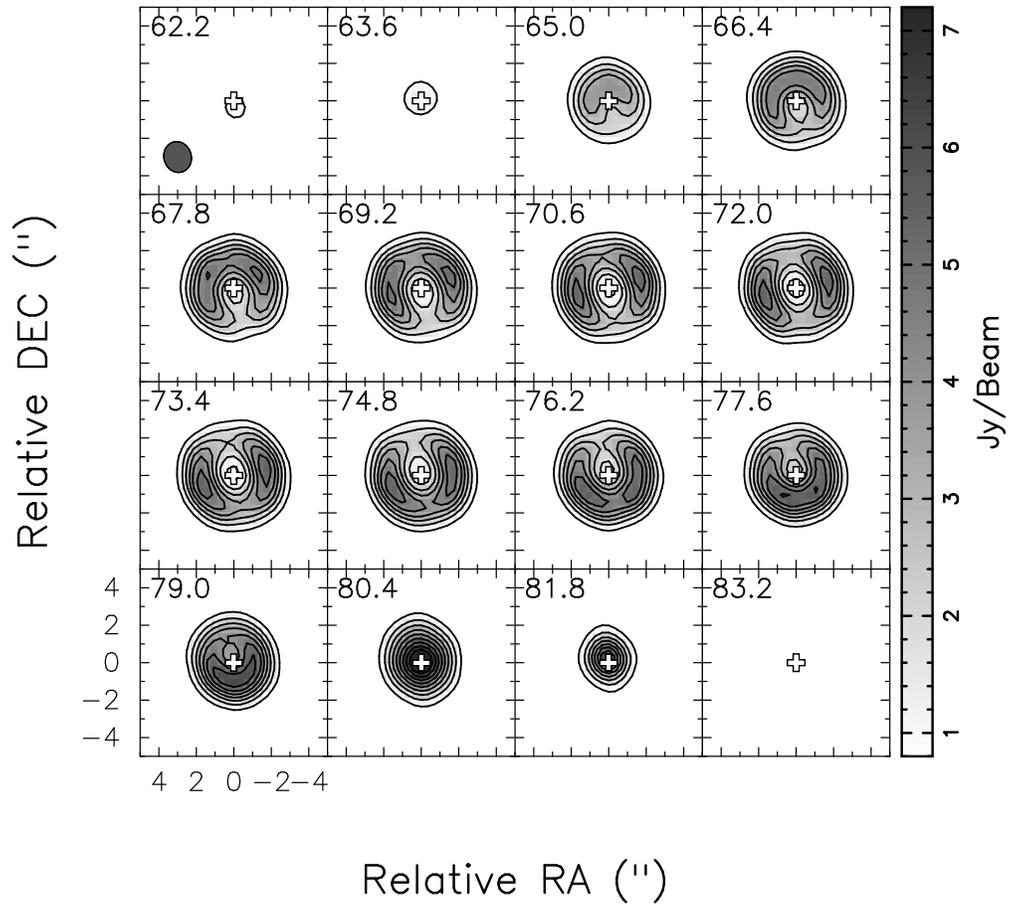}
\caption[]
{{The channel maps derived from our model. The contour levels, resolution,
velocity width and other map information are the same as those in Figure
\ref{obs_cha}.}\label{model_cha}}
\end{center}
\end{figure}

\begin{figure}[h]
\begin{center}
\includegraphics[width=120mm,angle=-90]{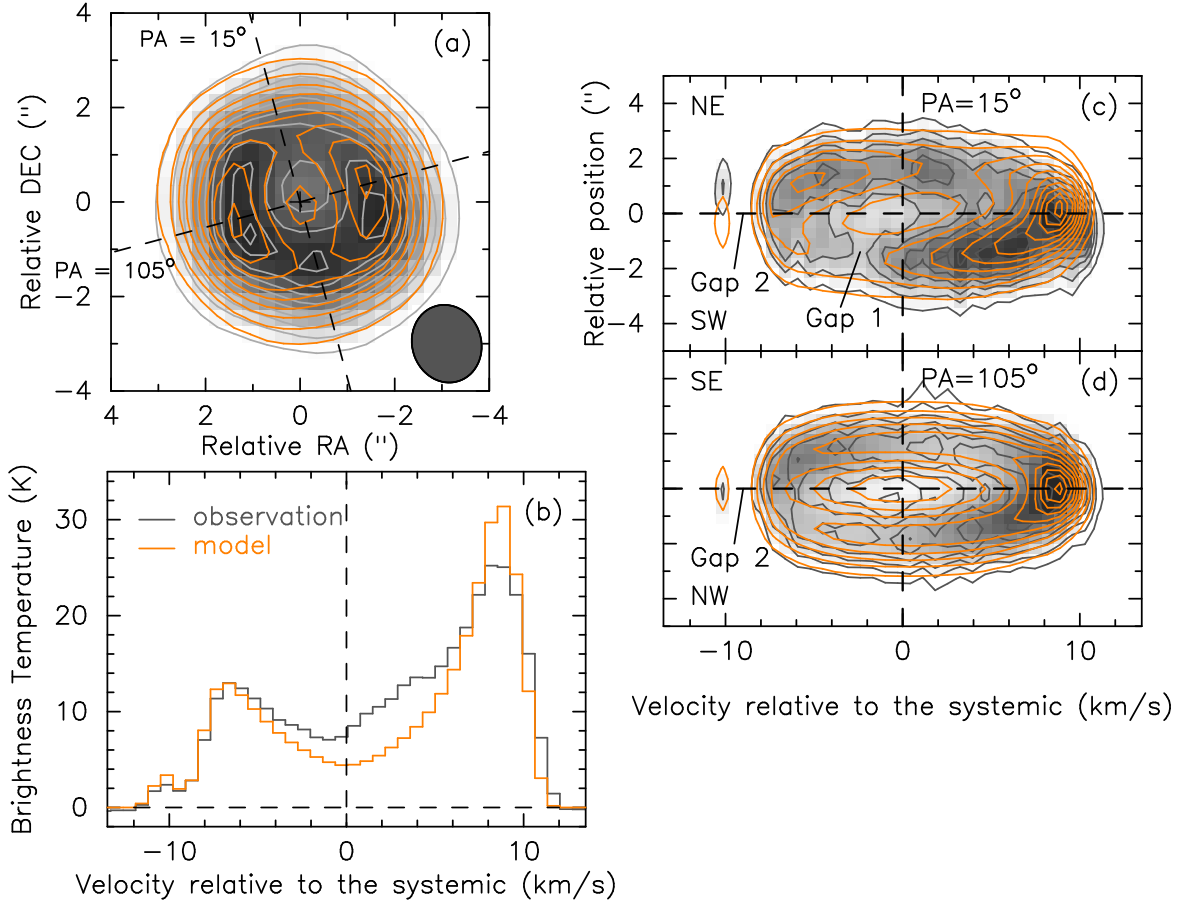}
\caption[]
{{Comparison of our model and the observation in the integrated intensity map, spectrum, and PV diagrams. Grey contours and spectrum are from the observation. Orange contours and spectrum are from our model. (a) shows the integrated intensity maps. The contour levels are the same as those in Figure \ref{co32_nearIR_midIR_co10}. Note that however, we do not include here the last contour of Figure \ref{co32_nearIR_midIR_co10}. (b) shows the spectra toward the central star position averaged over  a circular region with a diameter of \arcs{1}. (c) and (d) show the PV diagrams cut along PA $=$ 15$^\circ$ and PA $=$ 105$^\circ$, respectively. The contour levels are the same as those in Figure \ref{obs_pv}.}\label{model}}
\end{center}
\end{figure}

As shown in Figures \ref{model_cha} and \ref{model}, our model can reproduce
most of the observed structures and kinematics of the CO envelope. For
example, how the morphology and intensity of the emission change with
velocity in the channel maps, the equatorial enhanced emission in the
integrated intensity map, the intensity ratio of high red-shifted to high
blue-shifted peaks in the spectrum, and the emission gaps in the
blue-shifted part of the PV-diagrams. The mean optical depth of the CO J=3-2
emission estimated from the intensity ratio of the two peaks in our model
spectrum toward the source position is $\sim$1.

In our models, the values of $\Rin$ and $\Delta a$ are very similar to those
estimated by \citet{ueta 2005} from the near-IR image, which are \arcs{1.4}
and \arcs{0.8}, respectively. Also, $\epsilon$ $\sim$ 0.2, almost the same
as the value of 0.21 estimated from the optical image with the filter of
F547M \citep{ueta 2000}. The mean expansion velocity of the superwind is
also the same as the expansion velocity estimated by M2004 (10.5 \vkm). In
our model, $\Vturb$ $\sim$ 1 \vkm{}, and it is much important than $\Vth$,
which has a maximum value of $\sim$ 0.2 \vkm{}. Our PA is $\sim$
15$^\circ\pm5^\circ$,
in between those of \cite{ueta 2005} and N2009. The inclination angle is
estimated to be $\sim$ 25 $\pm $ 5$^\circ$, in between those estimated by
\cite{ueta 2005} and N2009. $\eqMlos$ obtained from our model is the same as
that in M2004. In our model for CO J=3-2, the superwind has a $\beta$ of
-0.8, much shallower than -2.0 found in CO J=1-0 (M2004), but steeper than
-0.4 found in IR dust emission \citep{cf 1991,meixner 1997}. 
On the other hand, in our model for CO J=3-2, the AGB wind has a $\gamma$ of -1.5, much steeper than -0.25 found in CO J=1-0 (M2004). We will discuss the differences later in the Discussion Section.
 
We can then derive 5 other quantities from our model: dynamical time at the
inner boundary of the superwind ($\tdyn$), duration of the superwind
($\tsw$), averaged mass-loss rate of the superwind ($\swMlos$) averaged over
$\theta$, as well as duration ($\tagb$) and mass-loss rate ($\agbMlos$) of
the AGB wind. We divide $\Rin$ by $\Va$ to obtain $\tdyn$ of $\sim$1590
years, divide $\Delta a$ by $\Va$ to obtain $\tsw$ of $\sim$1100 years. Our
$\swMlos$ is 1.8 $\times\ 10^{-5}\ \solarM$ yr$^{-1}$, which is lower by
40\% than that of M2004. We divide the thickness of the AGB wind
($\sim$\arcs{5.5}) by $\Va$ to obtain $\tagb$ of $\sim$ 6760 years. Our
$\tdyn$ is slightly larger than that obtained by M2004 (1240 years), with a
larger inner radius than theirs. Our $\tsw$ and $\tagb$ are similar to those
estimated by M2004, which are 840 and 6570 years, respectively. We
multiply $\eqMlos$ by $F$ to obtain $\agbMlos$ of 3.6 ($\pm$ 1.5) $\times\
10^{-6}\ \solarM$ yr$^{-1}$. $\agbMlos$ derived by M2004 is 5.1$\times\
10^{-6}\ \solarM$ yr$^{-1}$ and it equals our upper limit. In our model,
the CO emission of the AGB wind is pretty weak, with less than 3$\sigma$
even without having applied the observed \textsl{uv} coverage, and thus will
not be detected in our observation.

On the other hand, our model can not reproduce the following observed features:
\begin{enumerate}
\item{Our model is symmetric and thus can not reproduce the different radii between the eastern and western peaks in the observed integrated intensity map. The reason for this asymmetric is unclear, and will be discussed later.
}

\item{
The intensity excess at low red-shifted velocity in the observed spectrum [Figure \ref{model} (b)]. The spectrum is obtained by averaging over a circular region of \arcs{1} in diameter around the center, showing mainly the emission from the front and back walls of the envelope. Thus, the intensity excess indicates that there could be an additional material with low expansion velocity in the back wall (red-shifted part) of this envelope.
}

\item{
In our model, the highest blue-shifted emission is in the southwest and highest red-shifted emission is in the northeast of the center (Figure \ref{model_cha}), in opposite to those seen in the observation. The superwind could have a non-radial velocity component, for example, a poloidal velocity (directed from the equator to the poles). In the front wall of the superwind envelope, a poloidal velocity could increase the projected velocity of the north-eastern gas and decrease the projected velocity of the south-western gas near the map center. Same thing is for the back wall of the superwind envelope but with an opposite velocity sense. 
Alternatively, there could be an additional high-velocity component not included in our model.}
\end{enumerate}

\section{Discussion}

    \subsection{Superwind material in IR and CO observations}  

As compared to N2009, the superwind is better resolved in our map optimized
at a higher angular resolution. As a result, we can now study the superwind
in more details. At a higher rotational transition, CO J=3-2 can trace
warmer and denser material than CO J=1-0. Therefore, our CO J=3-2 map can
reveal the superwind material better than the CO J=1-0 map \citep{meixner
2004}, which shows both the superwind and AGB wind materials. Moreover,
as mentioned
before in Section 3, the two emission peaks in the equator in our CO
J=3-2 map are located slightly further away from the source than those in
the mid-IR map. It is likely because that the mid-IR emission mainly traces
the warmest material in the innermost part of the superwind, CO J=3-2
emission mainly traces the warm material in middle part of the superwind,
and CO J=1-0 emission traces the cool material further out in the superwind
and the AGB wind as well. 
Thus, in the superwind, the different
temperature power-law index $\beta$ in different tracers and
transitions may suggest that the power-law index becomes steeper with the
distance from the source. On the other hand, the temperature power-law index
of our AGB wind ($\gamma=-1.5$) in CO J=3-2 is in between those of the
superwind (-2.0) and AGB wind (-0.25) in CO J=1-0 \citep{meixner 2004}. This
is probably because that the superwind and AGB wind in the CO J=1-0
observation can not be clearly separated at low resolution. It is also possible that
our observation is not sensitive to the AGB wind.

    \subsection{Comparison with N2009 results}   

N2009 has proposed a model to compare with this observation. Their model
contains an inner toroidal superwind and an outer spheroidal AGB wind.
Figure \ref{model_sketch} right shows their model on top of our model. It is clear
from the figure that their torus could actually be considered as the dense
equatorial part of our ellipsoidal superwind. Note that the radial extent of
their torus is actually about one time larger than our superwind, and the PA
derived by our model is between that derived by N2009 and that estimated by
\cite{ueta 2005}. In their model, there was no radiative transfer and they
convolved their channel maps with a circular beam rather than the observed
elliptical beam. Their model reproduced the U- and inverted U-shaped
morphologies in the channel maps and the ring-like PV structure in the
PV-diagram. However, in order to reproduce the two elongated and clumpy
structures in the equator around the systemic velocity, their model needed a
distorted torus. Moreover, unlike that seen in the observation, the emission
intensity in their channel maps did not gradually increase from the
blue-shifted to red-shifted channels, but had a sudden increase in some
blue-shifted channels. In their PV-diagrams, Gap 2 was not seen on the high
blue-shifted velocity side. Our model, on the other hand, can reproduce all
the above observed features reasonably well. Furthermore, $\eqMlos$
estimated from our model is the same as that in M2004 and is in the typical
range for an AGB star. In contrast, N2009, by assuming that CO J=3-2 line is
optically thin, estimated a lower-limit mass-loss rate of $\sim$ $10^{-9}$
$\solarM$ yr$^{-1}$, which is 4 orders of magnitude lower than our $\eqMlos$
derived from our radiative transfer model. Therefore, radiative transfer is
really needed to estimate the mass-loss rate properly. Note that,
however, it is unclear how N2009 obtained such a low mass-loss rate since
the optical depth toward the central position in our model is only $\sim$ 1.

    \subsection{Shaping mechanism of elliptical PPN}
As mentioned in the Introduction Section, most of the low- to
intermediate-mass stars experience a structural change from spherical to
elliptical or to bipolar structure when they evolve from AGB to PPN phase
\citep{ueta 2000,sahai 2007}. It is still uncertain what
mechanism can cause this structural change. Moreover, most of PPNe contain
an additional density enhancement (torus-like) structure that is
perpendicular to their elliptical or bipolar structure.  I07134 is one of
the typical elliptical PPNe. It has an elliptical shell with an equatorial
density enhancement embedded in a round halo, indicating that it also
experiences the structural change mentioned above. At the present day, one
of the popular models to explain the forming mechanism of the density
enhancement is a model with a binary system. In such a model, if the
separation of the binary, wind velocity ejecting from the primary star and
the mass of the companion are in appropriate range, the material ejecting
from the primary star can accumulate around the equator with higher
expansion velocity than the poles, forming an asymmetric density enhancement
in the equator as seen in the observation \citep{mastrodemos 1999}. However,
in this scenario, the elliptical shell will appear oblate with the long axes
lying in the orbital plane, in contrast to that seen in I07134.

The elliptical structure of I07134 suggests that the poles of the superwind
have a higher expansion velocity than the equator. Thus if the binary system
is to work, an additional element will be needed to increase the expansion
velocity at the poles. A bipolar jet launched by accretion effect of the
companion \citep{mastrodemos 1998,frank 2004} is a candidate that can
produce high expansion velocity at the poles and it has been used to explain
the shaping mechanism of bipolar PPNe \cite[see, e.g.,][]{lee 2003}.
However, the observations toward I07134 do not show any high-velocity
emission. It is possible that the jet could be an atomic jet and thus could
not be detected in CO observation. Alternatively, a tenuous (and thus
unseen) isotropic post-AGB wind with higher expansion velocity than the
superwind could have been launched and interacted with the superwind,
producing the elliptical structure \citep{dwarkadas 1996} of I07134. In
either cases, I07134 could be in the transient phase changing from a round
to a bipolar structure.

However, the long-term (20 years) radial velocity observation of \cite{hrivnak 2011} seems to not support the binary scenario in I07134. If this is the case, the structural change of I07134 could be caused by some mechanisms due to a single star. \cite{DH 1996} proposed that a low rotation of a single AGB star would induce a preferential mass loss with higher velocities in the equatorial plane. However, in this scenario, the envelope would form an oblate shape, inconsistent with that seen in I07134. \cite{matt 2000} proposed the dipole magnetic field in an AGB star could lead the mass loss along the equator. But \cite{soker 2006} argued that the magnetic field would carry angular momentum away from the stellar envelope faster than the mass is lost by the wind, causing the star to spin down on a short timescale. In this case, the lifetime of magnetic field might not be long enough to produce the equatorial density enhancement structure. \cite{soker 1998,soker 2000} proposed that the concentration of cool magnetic spots toward the equator on the surface of an AGB star would lead to the equatorial density enhancement, because dust would form above the cool spots more than other area. Moreover, the mass loss velocity above the cool spots is lower than other directions due to the weak radiation from the cool spots, which has potential to form an elliptical envelope like I07134. However, it is unclear how above models can explain the asymmetric density distribution in the envelope of I07134.

\section{Conclusion}

We have reexamined and remodeled the SMA CO J=3-2 observational result of
I07134. Our main results are the following.

\begin{enumerate}

\item{
The CO map is consistent with the IR maps, showing an equatorial
density enhancement. The CO emission is located slightly further away from
the central source than the mid-IR emission, probably because that the
material is cooler in the outer part and thus better traced by the CO
emission.
}

\item{
Our model has two components, an inner ellipsoidal shell-like superwind with an equatorial density enhancement and an outer spheroidal AGB wind. The thick torus proposed by N2009 could actually be considered as the dense equatorial part of our ellipsoidal superwind.}

\item{
Our model can reproduce the observation reasonably well, better than the model proposed in N2009. The superwind mass-loss rate in the equator is estimated to be $\sim$ 3 $\times\ 10^{-5}\ \solarM$ yr$^{-1}$, the same as that derived by M2004 and is in the typical range for an AGB star. The mean expansion velocity of the superwind and AGB wind are 10.5 and 9.3 \vkm{}, respectively. The mass-loss durations of the superwind and AGB wind are 1100 and 6760 years, respectively, similar to those estimated in M2004. The superwind ended its ejection about 1590 years ago.}

\end{enumerate}

One of the popular models to explain the forming mechanism of the density
enhancement is a model with a binary system. However, in order for this
binary model to work for this PPN, an additional element, such as a bipolar
jet or a tenuous post-AGB wind, may be needed to produce the prolate
elliptical structure with the density enhancement in the equator.
Alternatively, the forming mechanism of the density enhancement could be due
to a single star.

We thank the referee for the valuable comments. We also thank Dr. Muthu
Mariappan, the PI of SMA 2004-32a, for the CO J=3-2 data of IRAS 07134+1005.
C.-H. Yang appreciates valuable discussion with Dr. Jun-Ichi Nakashima.
C.-F. Lee and C.-H. Yang acknowledge a grant from the National Science
Council of Taiwan (NSC96-2112-M-001-014-MY3).

\begin{deluxetable}{lc}
     \tablecaption{SMA observations of IRAS 07134+1005}
     \tablewidth{0pt}
     \tablehead{ 
    	   \colhead{Parameter} &  \colhead{Number}}
     \startdata
     R.A. (J2000.0)\tablenotemark{a}  & $\ra{07}{16}{10}{26}$  \\
     Dec1. (J2000.0)\tablenotemark{a}  & $\dec{+9}{59}{48}{0}$  \\
     Primary beam (FWHM) & $\sim$ \arcs{34} \\
     Velocity resolution ($\Delta V$) & $\sim$ 0.7 \vkm \\
     Flux calibrator & Callisto \\
     Passband calibrator  & Callisto \\
     gain calibrators & 0739+016, 0750+125 \\
     Synthesized beam  & \arcs{1.66}$\times$ \arcs{1.46} (PA = 17$^\circ$)  \\ 
     $\sigma$ ($\Delta V\ \sim$ 0.7 \vkm) &  0.27 \Jyb \\
     \enddata\label{tab:obs}
\tablenotetext{a}{The central star is assumed to be at this coordinate, as in \cite{meixner 2004}}.
\end{deluxetable}

\begin{deluxetable}{lcc}
\tablewidth{0pt}
\tablecaption{Model Parameters}
    \tablehead{\colhead{Parameter} &\colhead{Value} &\colhead{Reference}} \startdata 
    \multicolumn{3}{c}{Constant} \\ \tableline
     Distance & 2.4 kpc & 1 \\    
    CO/\H2 & 9.2 $\times 10^{-4}$ & 2 \\ 
    $\Rout$ & $\sim$ 19300 AU (\arcs{8}) & 2 \\  \tableline      
    \multicolumn{3}{c}{Free parameters} \\ \tableline        
    $\Rin$  &  \arcs{1.3} $\pm$ \arcs{0.1} ($\sim$3100 AU) \\
    $\Delta a$  &  \arcs{0.9} $\pm$ \arcs{0.1} ($\sim$2160 AU) \\    
    $\epsilon$=$1-\frac{a}{b}$  &  0.2 $\pm$ 0.05 \\
    PA & 15$^\circ \pm 5^\circ$ \\  
    $i$ & 25$^\circ \pm 5^\circ$ \\ 
    $\vsys$ & 72 $\pm $ 0.35 \vkm \\     
    $\Va$ &  9.3 $\pm $ 0.35 \vkm \\
    $\Vturb$  &  1 \vkm  \\    
    $\Tin$ & 70 $\pm$ 5 K \\
    $\FT$  &  0.25  $\pm$ 0.1 \\ 
    $\beta$ & -0.8 $\pm$ 0.2 \\
    $\gamma$ & -1.5 $\pm$ 0.5 \\
    $F$  &  0.12 $\pm$ 0.05 \\
    $\eqMlos$  &  3 ($\pm$ 1) $\times 10^{-5}\ \solarM$ yr$^{-1}$ \\    
    $\alpha$ & 0.8  $\pm$ 0.1  \\ \tableline 
    \multicolumn{3}{c}{Output} \\ \tableline
    $\tdyn$  & $\sim$ 1590 yr  \\
    $\tsw$  &  $\sim$ 1100 yr \\
    $\tagb$ &  $\sim$ 6760 yr  \\ 
    averaged $\swMlos$  &  1.8 ($\pm$ 0.6) $\times\ 10^{-5}\ \solarM$ yr$^{-1}$ \\    
    $\agbMlos$  &  3.6 ($\pm$ 1.5) $\times\ 10^{-6}\ \solarM$ yr$^{-1}$ \\                   
    \enddata\label{tab:mod}
    \tablerefs {(1) \cite{knapp 2000};  (2) \cite{meixner 2004}.}
\end{deluxetable}

\renewcommand{\baselinestretch}{1}
\renewcommand{\baselinestretch}{1.4} 

\renewcommand{\baselinestretch}{1}
\renewcommand{\baselinestretch}{1.4} 

\renewcommand{\baselinestretch}{1}
\renewcommand{\baselinestretch}{1.4}

\renewcommand{\baselinestretch}{1}
\renewcommand{\baselinestretch}{1.4}

\renewcommand{\baselinestretch}{1}
\renewcommand{\baselinestretch}{1.4}

\renewcommand{\baselinestretch}{1}
\renewcommand{\baselinestretch}{1.4}

\end{document}